\documentclass[pra,aps,twocolumn]{revtex4}
\input epsf.tex

\newcommand{\be}{\begin{equation}}
\newcommand{\ee}{\end{equation}}
\newcommand{\ba}{\begin{eqnarray}}
\newcommand{\ea}{\end{eqnarray}}
\newcommand{\ban}{\begin{eqnarray*}}
\newcommand{\ean}{\end{eqnarray*}}

\newcommand{\ket}[1]{\mbox{$ | #1 \rangle $}}
\newcommand{\bra}[1]{\mbox{$ \langle #1 | $}}

\newcommand{\myone}{\leavevmode\hbox{\small1\normalsize\kern-.33em1}}

\newcommand{\moy}[1]{\langle #1 \rangle}

\newcommand{\beq}{\begin{equation}}
\newcommand{\eeq}{\end{equation}}
\newcommand{\beqa}{\begin{eqnarray}}
\newcommand{\eeqa}{\end{eqnarray}}

\def\opone{\leavevmode\hbox{\small1\normalsize\kern-.33em1}}

\begin{document}

\title{Testing quantum correlations versus single-particle properties \\ within Leggett's model and beyond}
\author{Cyril Branciard$^1$}
\email{cyril.branciard@physics.unige.ch}
\author{Nicolas Brunner$^1$, Nicolas Gisin$^1$, \\ Christian Kurtsiefer$^2$, Antia Lamas-Linares$^2$, Alexander Ling$^2$}
\author{Valerio Scarani$^2$}
\affiliation{$^1$Group of Applied Physics, University of Geneva,
Geneva, Switzerland \\ $^2$Centre for Quantum Technologies / Physics
Dept., National University of Singapore, Singapore}
\date{\today}

\maketitle

{\bf

Quantum theory predicts and experiments confirm that nature can
produce correlations between distant events that are nonlocal in the
sense of violating a Bell inequality \cite{bell64}. Nevertheless,
Bell's strong sentence {\it Correlations cry out for explanations}
remains relevant. The maturing of quantum information science and
the discovery of the power of nonlocal correlations, e.g. for
cryptographic key distribution beyond the standard Quantum Key
Distribution schemes
\cite{crypto_beyond_QKD_1,crypto_beyond_QKD_2,devindep}, strengthen
Bell's wish and make it even more timely.

In 2003, Leggett proposed an alternative model for nonlocal
correlations \cite{leg03}, that he proved to be incompatible with
quantum predictions. We present here a new approach to this model,
along with new inequalities for testing it. Remarkably these
inequalities can be derived in a very simple way, assuming only the
non-negativity of probability distributions; they are also stronger
than previously published Leggett-type inequalities
\cite{leg03,Nature_Vienna,PRL_Vienna,PRL_Singap}. The simplest of
these inequalities is experimentally violated. Then we go beyond
Leggett's model, and show that one cannot ascribe even partially
defined individual properties to the components of a maximally
entangled pair.

}

\bigskip

Formally, a correlation is a conditional probability distribution
$P(\alpha,\beta|\vec a,\vec b)$, where $\alpha,\beta$ are the
outcomes observed by two partners, Alice and Bob, when they perform
measurements labeled by $\vec a$ and $\vec b$, respectively. On the
abstract level, $\vec a$ and $\vec b$ are merely inputs, freely and
independently chosen by Alice and Bob. On a more physical level,
Alice and Bob hold two subsystems of a quantum state; in the simple
case of qubits, the inputs are naturally characterized by vectors on
the Poincar\'e sphere, hence the notation $\vec a,\vec b$.

How should one understand nonlocal correlations, in particular those
corresponding to entangled quantum states? A natural approach
consists in decomposing $P(\alpha,\beta|\vec a,\vec b)$ into a
statistical mixture of hopefully simpler correlations: \be
P(\alpha,\beta|\vec{a},\vec{b}) = \int \mathrm{d} \lambda
\rho(\lambda) \ P_\lambda(\alpha,\beta|\vec{a},\vec{b}) \ .
\label{Pmixture}\ee Bell's locality assumption is
$P_\lambda(\alpha,\beta|\vec{a},\vec{b})=
P_\lambda(\alpha|\vec{a})P_\lambda(\beta|\vec{b})$; admittedly the
simplest choice, but an inadequate one as it turns out: quantum
correlations violate Bell's locality \cite{bell64}. Setting out to
explore other choices, it is natural to require first that the
$P_\lambda$'s fulfill the so-called {\it no-signaling condition},
i.e., that none of the correlations $P_\lambda$ results from a
communication between Alice and Bob. This can be guaranteed by
ensuring space-like separation between Alice and Bob. Non-signaling
correlations happen without any time-ordering: there is not a first
event, let's say at Alice's side, that causes the second event via
some {\it spooky action at a distance}. One may phrase it
differently: non-signaling correlations happen from outside
space-time, in the sense that there is no story in space-time that
tells us how they happen. This is the case in orthodox quantum
physics, or in some illuminating toy models like the nonlocal box of
Popescu and Rohrlich (PR-box) \cite{PR}. Mathematically, the
no-signaling condition reads
$P_\lambda(\alpha|\vec{a},\vec{b})=P_\lambda(\alpha|\vec{a})$ and
$P_\lambda(\beta|\vec{a},\vec{b})=P_\lambda(\beta|\vec{b})$: the
local statistics of Alice are not influenced by Bob's choice of
measurement, and reciprocally.

In 2003, Leggett proposed another model of the form
(\ref{Pmixture}), which can also be experimentally tested against
quantum predictions \cite{leg03}. This model was recently brought
into focus by the work of Gr\"oblacher {\it et al.}
\cite{Nature_Vienna}. The basic assumption of Leggett's model is
that locally everything happens as if each single quantum system
would always be in a pure state. We shall be concerned here with the
case of binary outcomes $\alpha,\beta=\pm1$, though generalizations
are possible. In this case, the supplementary variables $\lambda$ in
Leggett's model describe pure product states of two qubits, denoted
by normalized vectors $\vec u, \vec v$ on the Poincar\'e sphere:
\beq \label{lambda_leggett} \lambda=\ket{\vec u}\otimes\ket{\vec v}
\ , \eeq and the local expectation values have the usual form as
predicted by quantum physics: \beqa \label{MA}
\moy{\alpha}_{\lambda} & = & \bra{\vec
u} \, \vec a \cdot \vec\sigma \, \ket{\vec u}=\vec u\cdot\vec a \ , \\
\label{MB} \moy{\beta}_{\lambda} & = &\bra{\vec v} \, \vec b \cdot
\vec\sigma \, \ket{\vec v}=\vec v\cdot\vec b \ . \eeqa If the qubits
are encoded in the polarization of photons, as in Leggett's initial
idea, then the assumption is that each photon should locally behave
as if it were perfectly polarized (in the directions $\vec u$ and
$\vec v$), and the local observations, conditioned on each
$\lambda$, should fulfill Malus' law. It is worth emphasizing that
Leggett's assumption concerns exclusively the local marginals
$\moy{\alpha}_{\lambda}$ and $\moy{\beta}_{\lambda}$ of the
probability distributions $P_\lambda$, while nothing is specified
about the correlation coefficients $\moy{\alpha\beta}_{\lambda}$.
Leggett's model can thus still be nonlocal, and can in general
violate a Bell inequality.

Both in the original paper \cite{leg03} and in \cite{Nature_Vienna},
the model was presented by implicitly assuming a time-ordering of
the events. Any model based on such an assumption had already been
falsified by the so-called before-before experiment
\cite{before_before_1,before_before_2}, as Suarez emphatically
stressed \cite{suarez}. However, assumptions (\ref{MA}--\ref{MB})
clearly define non-signaling correlations, and Leggett's model can
be defined without any reference to time-ordering. As a consequence,
its study does add something to our understanding of quantum
non-locality. But what exactly? In what are such $P_{\lambda}$'s
``simpler'' than the usual quantum correlations? To answer these
questions, we recall that, in quantum theory, the singlet state is
such that the properties of the pair are sharply defined (the state
is pure), but the properties of the individual qubits are not. In
this perspective, Leggett's model is an attempt of keeping the
correlations while reintroducing sharp properties at the individual
level as well.

Leggett's model cannot reproduce the correlations of the singlet
state. Experimental falsifications have already been reported, first
under additional assumptions \cite{Nature_Vienna}, then more
directly \cite{PRL_Vienna,PRL_Singap}. These works relied on the
violation of so-called \textit{Leggett-type inequalities}. Analog to
Bell's inequalities, these criteria say that, under Leggett's
assumptions (\ref{MA}) and (\ref{MB}), a measurable quantity $L$
should satisfy $L\leq L_{\max}$, while quantum theory predicts that
$L>L_{\max}$ can be observed for suitable measurements. An important
feature of Leggett-type inequalities is that, contrary to Bell's
inequalities, the bound $L_{max}$ is not a fixed number: instead,
like the model itself, it depends on the quantum measurements that
are performed. Consequently all experimental data aiming at
disproving Leggett's model should present evidence that the settings
used in the experiment have been properly adjusted.

All previously available derivations of Leggett-type inequalities
were quite lengthy and failed to suggest possible improvements or
generalizations. We have found a much more straightforward
derivation (see the Methods section), simply based on the fact that
each $P_\lambda$ must be a valid probability distribution, so in
particular $P_\lambda(\alpha,\beta|\vec{a},\vec{b})\geq 0$.
Remarkably, this constraint of \textit{non-negativity of
probabilities}, weak as it may seem, is enough to induce an
observable incompatibility between Leggett's model and quantum
predictions. In our derivation, it also appears that the previously
derived Leggett-type inequalities are sub-optimal; among the
improved inequalities that our approach suggests, the simplest one
reads \ba \frac{1}{3} \sum_{i=1}^3 |C(\vec a_i,\vec b_i) + C(\vec
a_i,\vec b'_i)| \nonumber \\ \equiv L_3(\varphi) & \leq & 2 -
\frac{2}{3} |\sin \frac{\varphi}{2}| \label{new_Leggett_ineq} \ea
where $C(\vec a,\vec b)=\sum_{\alpha,\beta}\alpha\beta
P(\alpha,\beta|\vec a ,\vec b)$ is the usual correlation
coefficient. This inequality holds provided the three measurements
on Alice's side and six on Bob's fulfill some relations; a possible
set of measurements is given in Figure~1. 
 For the singlet state, quantum mechanics predicts
$C_{\Psi^-}(\vec{a},\vec{b}) = - \vec{a} \cdot \vec{b}$. Thus, for
the settings just defined, $L_3(\varphi)$ is \ba L_{\Psi^-}(\varphi)
= 2 |\cos \frac{\varphi}{2}| \ . \label{L_QM} \ea This expression
violates inequality (\ref{new_Leggett_ineq}) for a large range of
values $\varphi$.

\begin{center}
\begin{figure}
\epsfxsize=7cm \epsfbox{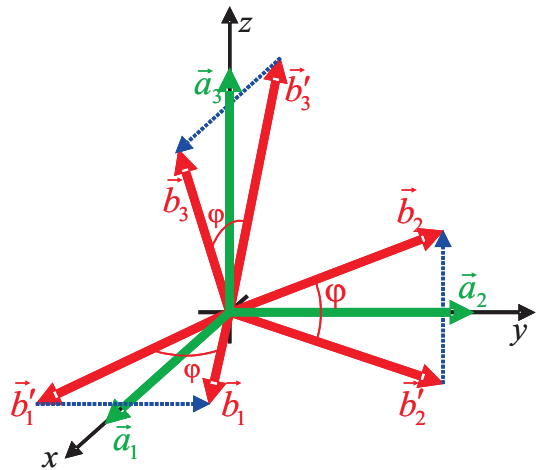} \caption{{\bf Alice's
(green) and Bob's (red) measurement settings used to test inequality
(\ref{new_Leggett_ineq}).} In order for inequality
(\ref{new_Leggett_ineq}) to hold, Bob's settings must be such that
the three pairs $(\vec b_i, \vec b_i')$ form the same angle
$\varphi$, and the three directions $\vec e_i$ of $\vec b_i - \vec
b_i'$ (blue) are orthogonal. The best violation is obtained when
Alice's settings $\vec a_i$ are chosen to be along the directions of
$\vec b_i + \vec b_i'$.} \label{fig_settings}
\end{figure}
\end{center}

\begin{figure}
\centerline{\epsfxsize=70mm \epsfbox{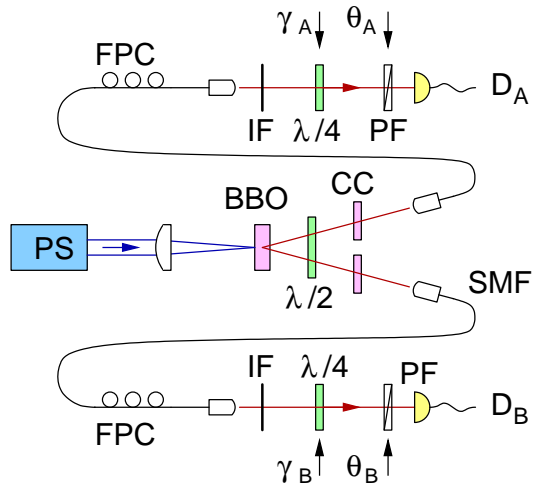}}
\caption{{\bf Experimental
  setup.} Polarization-entangled photon pairs are generated in a nonlinear
  optical crystal (BBO) and coupled into single mode optical fibers
  (SMF), similarly as in \cite{PRL_Singap}. Polarization measurements in
  arbitrary settings for each photon are performed with polarization filters
  (PF) and quarter wave plates ($\lambda/4$), followed by
  single photon detectors (D$_{A,B}$). More details can be found in
  Supplementary Information I.}
\label{fig_exp_setup}
\end{figure}

In order to test the Leggett-type inequality
(\ref{new_Leggett_ineq}) in an experiment, we prepared
polarization-entangled photon pairs into single mode optical fibers
in a close approximation to a singlet state, similarly as in
\cite{PRL_Singap}. In our setup (see
Figure~2), 
we choose the settings of
polarization measurements $\vec{a_i}, \vec{b_i}$ and $\vec{b_i}'$
for the individual photons by dialing in appropriate orientation
angles $\gamma_{A,B}$ for two quarter wave plates ($\lambda/4$), and
angles $\theta_{A,B}$ for two absorptive polarization filters (PF).
Details about the experimental implementation can be found in
Supplementary Information I. Through four consecutive coincidence
measurements between photodetectors $D_{A,B}$ for all combinations
of settings $\vec{a},-\vec{a}$ and $\vec{b},-\vec{b}$, we establish
an experimental value for a correlation coefficient
$C(\vec{a},\vec{b})$.

The correlation coefficients necessary to compose values for
$L_3(\varphi=\pm30^\circ)$ were obtained with an integration time of
$T=60$\,s per point, leading to values of $1.9068\pm0.0009$ for
$\varphi=-30^\circ$ and $1.9005\pm0.0010$ for $\varphi=30^\circ$.
 This corresponds to a violation of the
bound for $L_3(\varphi=-30^\circ)$ and $L_3(\varphi=+30^\circ)$ in
Leggett's model by 83.7 and 74.5 standard deviations, respectively.

The asymmetry in the measured values of $L_3(\varphi)$ is an
indication for experimental imperfections in the accuracy of the
settings, such as a possible misalignment of one of the quarter wave
plates with respect to the polarizing filters. To test this
alignment, we collected values for $L_3$ over a larger range of
$\varphi$ with an integration time of $T=15$\,s per setting
(Figure~3).
The variation of $L_3$ with
$\varphi$ is compatible with the quantum mechanical prediction for a
singlet state with residual colored noise and an orientation
uncertainty of the quarter wave plate of 0.2$^\circ$.

\begin{figure}
\centerline{\epsfxsize=78mm \epsfbox{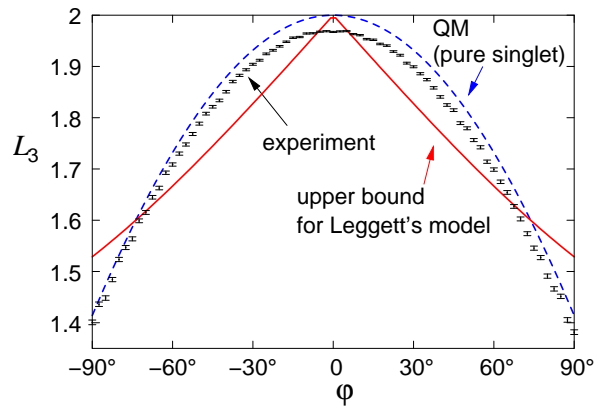}}
\caption{{\bf Violation of Leggett's model.} The experimental
  values for $L_3$ over a range of separation angles $\varphi$ (points with
  error bars) violate the bound given by Leggett's model (solid line), and follow
  qualitatively the quantum mechanical (QM) prediction (dashed line). The error bars indicate $\pm1$ standard deviations of propagated Poissonian
counting statistics assumed for photodetection events. The largest
violations of inequality (\ref{new_Leggett_ineq}) are found for
$\varphi=\pm25^\circ$ with 40.6 and 38.1 standard deviations,
respectively.} \label{fig_experimental_plot}
\end{figure}

The falsification of Leggett's model proves that it is impossible to
reconstruct quantum correlations from hypothetical, more elementary
correlations in which individual properties would be sharply
defined. Let us argue that a much stronger statement holds, namely,
that individual properties cannot even be partially defined.

We first consider the following straightforward generalization of
Leggett's model: we allow the ``local states" $\lambda$ to be mixed
states, e.g. photons with a degree of polarization $\eta$. So, we
replace (\ref{MA}) and (\ref{MB}) by
\ba \begin{array}{lcl}\moy{\alpha}_{\lambda} &=& \eta \ \vec{u} \cdot \vec{a} \ , \\
\moy{\beta}_{\lambda} &=& \eta \ \vec{v} \cdot \vec{b} \
,\end{array}& \mathrm{with} & 0 < \eta \leq 1 \ . \ea The derivation
of inequalities for this model follows exactly the same pattern as
for usual Leggett-type inequalities. In particular, the generalized
version of (\ref{new_Leggett_ineq}) is \ba L_3(\varphi) & \leq & 2 -
\frac{2}{3} \, \eta \, |\sin \frac{\varphi}{2}|
\label{ineg_eta_leggett3} \ea which, for angles $\varphi$ small
enough, is violated by $L_{\Psi^-}(\varphi)$ for any value of $\eta
> 0$. Thus, as soon as the degree of purity of the ``local states"
is non-zero, this generalization of Leggett's model also fails to
reproduce quantum mechanical predictions (see Supplementary
Information II for a more complete analysis of this generalization).

Experimentally, one cannot expect to conduct a meaningful comparison
between those two predictions down to $\eta=0$, due to imperfections
in the state preparation. From the measurement of Figure~3,
however, we can claim experimental evidence of a violation for all
$\eta \geq 0.56$, with a statistical significance of at least 3.65
standard deviations, thus putting a lower bound for this class of
models (see Figure~4). 

\begin{figure}
\centerline{\epsfxsize=78mm \epsfbox{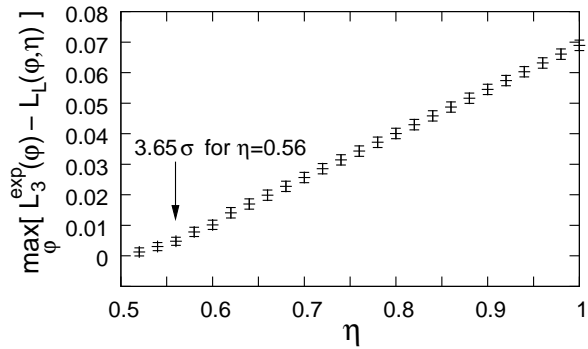}} \caption{{\bf
Experimental test of the generalized Leggett inequality for
  different degrees of purity $\eta$.}
  From the measurement shown in Figure~3, 
we extract the maximal excess of the experimental $L_3(\varphi)$
with respect to the upper bound (denoted $L_L$) in inequality
(\ref{ineg_eta_leggett3}), for various degrees of purity $\eta$. For
$\eta<0.6$, data at $\varphi=-15^\circ$ are considered, while for
$\eta>0.6$, the largest violation are observed for
$\varphi=-25^\circ$.
} \label{fig_etalimit}
\end{figure}

It is then natural to conjecture that no model of form
(\ref{Pmixture}), with non-signaling correlations  $P_\lambda$, can
perfectly reproduce the correlations of the singlet state
$C_{\Psi^-}(\vec{a},\vec b) = - \vec a \cdot \vec b$, unless \ba
\moy{\alpha}_{\lambda}= \moy{\beta}_{\lambda}&=&0 \ea for all
measurements $\vec{a}$ and $\vec b$ (except perhaps for a few
$\lambda$'s in a zero-measure set). In the Methods section we prove
this conjecture for models with discrete supplementary variables
$\lambda$; continuity arguments should allow to extend the result to
more general models. We thus have a necessary condition for a
non-signaling model to be compatible with quantum mechanics.
However, given a model with non-trivial marginals, finding an
explicit inequality that can be tested experimentally against
quantum predictions is another problem; for that, one needs the
specific details of the model.

\bigskip

In summary, with the general goal of improving our understanding of
quantum correlations, we reformulated Leggett's model. No
time-ordering of the events was assumed, and all assumptions were
made on the local part of the correlations. We derived new
Leggett-type inequalities, simpler and stronger than previously
known. The simplest version of our inequalities has been
experimentally violated. Finally we investigated more general models
\textit{\`a la Leggett}, for which we only imposed the no-signaling
condition. We argued that any such model with biased marginals is
incompatible with quantum predictions for the singlet state. This
shows that quantum correlations cannot be reconstructed from
``simpler'' correlations in which the individual properties would be
even partially defined. Nature is really such that, in some cases,
individual properties are completely lost while global properties
are sharply defined.

Our result is in good agreement with the recent work of Colbeck and
Renner \cite{ColRen}, who have derived general inequalities to
falsify such models with non trivial marginals. An example of a
non-signaling model that successfully reproduces the singlet
correlations can be found in \cite{simulPRbox}; indeed this model
has unbiased marginals. It is also worth mentioning the nonlocal
model of Toner and Bacon \cite{Toner}, which reproduces analytically
the singlet correlations with one bit of communication. In this
model, the probability distributions $P_\lambda$ have non vanishing
marginals; however, the $P_\lambda$'s are signaling. The remarkable
property of the Toner-Bacon model is that the communication is
cleverly hidden, such that the final probability distribution $P$ is
non-signaling.

This work is part of the general research program that looks for
nonlocal models compatible or incompatible with quantum predictions.
The goal is to find out what is essential in quantum correlations.
Here we found that in order to simulate or to decompose the singlet
correlations, one can't use non-signaling correlations $P_\lambda$
with non-trivial marginals. This nicely complements the result of
\cite{BrunnerNJP}, that the correlations corresponding to very
partially entangled states, hence large marginals, can't be
simulated by a single PR-box, which has trivial marginals.

\section*{METHODS}

\subsection{Simple derivation of inequalities \\ that test Leggett's model}
\label{appderiv}

\subsubsection{Convenient notations}

As mentioned, in this paper we focus on the case of binary outcomes
$\alpha,\beta=\pm1$. In this case, the correlations can conveniently
be written as \beqa \label{Pab} && P_\lambda(\alpha,\beta|\vec
a,\vec b)=\\ \nonumber && \quad \frac{1}{4}\left(1+\alpha \
M_\lambda^A(\vec a,\vec b)+\beta \ M_\lambda^B(\vec a,\vec
b)+\alpha\beta \ C_\lambda(\vec a,\vec b)\right) . \eeqa This
expression allows one to clearly distinguish the marginals
$M_\lambda^A(\vec a,\vec b)=\sum_{\alpha,\beta} \alpha
P_\lambda(\alpha,\beta|\vec a,\vec b)$ on Alice's side and
$M_\lambda^B(\vec a,\vec b)=\sum_{\alpha,\beta} \beta
P_\lambda(\alpha,\beta|\vec a,\vec b)$ on Bob's, and the correlation
coefficient $C_\lambda(\vec a,\vec b)=\sum_{\alpha,\beta}
\alpha\beta P_\lambda(\alpha,\beta|\vec a,\vec b).$ Throughout the
Methods section, we shall use these notations; in the main text, we
have used more standard and simplified notations, the correspondence
being $\moy{\alpha}_{\lambda}=M_\lambda^A(\vec a,\vec b)$,
$\moy{\beta}_{\lambda}=M_\lambda^B(\vec a,\vec b)$. The no-signaling
condition is $M_\lambda^A(\vec a,\vec b)=M_\lambda^A(\vec a)$ and
$M_\lambda^B(\vec a,\vec b)=M_\lambda^B(\vec b)$.

In order for the decomposition (\ref{Pmixture}) to be a valid
mixture of correlations, all distributions $P_\lambda$ should be
non-negative. As we said, this constraint is enough to derive
Leggett-type inequalities. From eq.~(\ref{Pab}), one can see that
the non-negativity implies the general constraints: \be
|M_\lambda^A(\vec{a}) \pm M_\lambda^B(\vec{b})| \leq 1 \pm
C_\lambda(\vec{a},\vec{b}) \ . \label{constr_non_neg} \ee
Constraints on the marginals $M_\lambda^A$ or $M_\lambda^B$ thus
imply constraints on the correlation coefficients $C_\lambda$, and
{\it vice versa}.

Let's now consider one measurement setting $\vec{a}$ for Alice and
two measurement settings $\vec b, \vec b'$ for Bob, and let's
combine the previous inequalities (\ref{constr_non_neg}) that we get
for $(\vec{a},\vec{b})$ and $(\vec{a},\vec{b}')$. Using the triangle
inequality, one gets \be |C_\lambda(\vec{a},\vec b) \pm
C_\lambda(\vec{a},\vec b')| \leq 2 - |M_\lambda^B(\vec b) \mp
M_\lambda^B(\vec b')| \ . \ee

These constraints must hold for all probability distributions
$P_\lambda$. After integration over the $\lambda$'s, one gets, for
the averaged correlation coefficients $C(\vec{a},\vec{b}) = \int
\mathrm{d} \lambda \rho(\lambda) C_\lambda(\vec{a},\vec{b})$ \ba
|C(\vec{a},\vec b) \pm C(\vec{a},\vec b')| \leq 2 - \int \mathrm{d}
\lambda \rho(\lambda) \ |M_\lambda^B(\vec b) \mp M_\lambda^B(\vec
b')| \ . \nonumber \\ \label{basic_ineq} \ea This inequality is
general for all models with ``local marginals", i.e. that fulfil the
no-signaling condition.

\subsubsection{Derivation of a simple Leggett-type inequality}
\label{section_deriv_leggett}

Now we derive an inequality satisfied by Leggett's specific model,
that can be experimentally tested. Inequality (\ref{basic_ineq})
implies, for the particular form of eq. (\ref{MB}) for Bob's
marginals: \be |C(\vec{a},\vec b) + C(\vec{a},\vec b')| \leq 2 -
\int \mathrm{d} \lambda \rho(\lambda) \ |\vec v \cdot (\vec b - \vec
b')| \ . \label{basic_ineq_leggett} \ee

Let's consider three triplets of settings $(\vec{a}_i, \vec b_i,
\vec b'_i)$, with the same angle $\varphi$ between all pairs ($\vec
b_i, \vec b'_i$), and such that $\vec b_i - \vec b'_i = 2 \sin
\frac{\varphi}{2} \vec{e}_i$, where
$\{\vec{e}_1,\vec{e}_2,\vec{e}_3\}$ form an orthogonal basis (see
Figure~1). 
After
combining the three corresponding inequalities
(\ref{basic_ineq_leggett}), using the fact that $\sum_{i=1}^3
|\vec{v} \cdot \vec{e}_i| \geq 1$ and the normalization $\int
\mathrm{d} \lambda \rho(\lambda) = 1$, we finally get the
Leggett-type inequality (\ref{new_Leggett_ineq}).

For a pure singlet state, inequality (\ref{new_Leggett_ineq}) is
violated when $|\varphi| < 4 \arctan\frac{1}{3} \simeq 73.7^\circ$,
and the maximal violation is obtained for $|\varphi| = 2
\arctan\frac{1}{3} \simeq 36.9^\circ$. In the case of imperfect
interference visibility $V$ ($\tilde{L}_{\Psi^-}(\varphi) = 2 V
|\cos \frac{\varphi}{2}|$), a violation can still be observed as
long as $V > V_{th}^{(3)} = \sqrt{1-(\frac{1}{3})^2} =
\frac{2\sqrt{2}}{3} \simeq 94.3 \%$.

Note that other Leggett-type inequalities can be derived, as we
mention in Supplementary Information III.

\subsection{Any non-signaling model \\ must have vanishing
marginals} \label{section_generalization}

We prove here that all marginals in a non-signaling model must
necessarily satisfy the constraints (\ref{constr_gen_1}) and
(\ref{constr_gen_2}) below; and we argue that this in turn implies
the claim made in the main text, namely: all the marginals must
vanish (apart perhaps on a zero-measure subset of $\lambda$'s).

In order for a general non-signaling model to reproduce all quantum
correlations $C_{\Psi^-}(\vec{a},\vec b) = - \vec a \cdot \vec b$ of
the singlet state, one must have, according to equation
(\ref{basic_ineq}), for all $\vec a, \vec b, \vec b'$ on the
Poincar\'e sphere: \ba \int \mathrm{d} \lambda \rho(\lambda) \
|M_\lambda^B(\vec b) \pm M_\lambda^B(\vec b')| \leq 2 - |\vec a
\cdot (\vec b \mp \vec b')| \ea and therefore, for all $\vec b, \vec
b'$: \ba \int \mathrm{d} \lambda \rho(\lambda) \ |M_\lambda^B(\vec
b) \pm M_\lambda^B(\vec b')| \leq 2 - ||\vec b \mp \vec b'||
\label{constr_gen}\ea where $||\cdot||$ is the euclidian norm.

In the case where $\vec b' = - \vec b$, the first constraint of eq.
(\ref{constr_gen}) implies that for all $\vec b$ \ba \int \mathrm{d}
\lambda \rho(\lambda) \ |M_\lambda^B(\vec b) + M_\lambda^B(-\vec b)|
= 0 \ . \label{constr_gen_1} \ea In the case $\vec b' \to \vec b$,
the two vectors being normalized, we have $2 - ||\vec b + \vec b'||
= 2 - \sqrt{4-||\vec b - \vec b'||^2} \sim \frac{||\vec b - \vec
b'||^2}{4}$ and therefore the second constraint of
(\ref{constr_gen}) implies that for all $\vec b$ \ba \lim_{\vec b'
\to \vec b} \int \mathrm{d} \lambda \rho(\lambda) \
\frac{|M_\lambda^B(\vec b) - M_\lambda^B(\vec b')|}{||\vec b - \vec
b'||} = 0 \ . \label{constr_gen_2} \ea

Let us now prove, in the case of discrete $\lambda$'s ($\lambda \in
\{\lambda_i\}$), that (\ref{constr_gen_1}) and (\ref{constr_gen_2})
in turn necessarily imply that the marginals $M_\lambda^B(\vec b)$
must vanish. In this case, the integral $\int \mathrm{d} \lambda
\rho(\lambda)$ should be changed to a discrete sum $\sum_i
p_{\lambda_i}$.

Constraint (\ref{constr_gen_1}) implies indeed that for all
$\lambda_i$ (such that $p_{\lambda_i} > 0$) and for all $\vec b$,
$M_{\lambda_i}^B(-\vec b) = -M_{\lambda_i}^B(\vec b)$, i.e.
$M_{\lambda_i}^B$ must be an odd function of $\vec b$. This is a
very natural property that one would impose to such a model; in
particular, Leggett's model has indeed odd marginals.

In addition, constraint (\ref{constr_gen_2}) implies that for all
$\lambda_i$ and for all $\vec b$, $\lim_{\vec b' \to \vec b}
\frac{|M_{\lambda_i}^B(\vec b) - M_{\lambda_i}^B(\vec b')|}{||\vec b
- \vec b'||} = 0$, i.e. that all $M_{\lambda_i}^B$ are
differentiable, and their derivative is 0 for all $\vec b$:
therefore they are constant. Since they have to be odd functions,
then necessarily they are equal to zero.

In conclusion, for discrete $\lambda$'s, Bob's marginals
$M_{\lambda_i}^B(\vec b)$ must all vanish; of course, the same
reasoning holds for Alice's marginals $M_{\lambda_i}^A(\vec a)$.
This result should also be valid for any distribution
$\rho(\lambda)$, at least those physically reasonable (e.g.
piecewise continuous): we conjecture that for any reasonable model
to reproduce the quantum correlations of the singlet state,
necessarily the marginals must vanish, in the sense that \ba &&
\forall \vec a, \quad \int
\mathrm{d} \lambda \rho(\lambda) |M_{\lambda}^A(\vec a)| = 0 \ , \\
&& \forall \vec b, \quad \int \mathrm{d} \lambda \rho(\lambda)
|M_{\lambda}^B(\vec b)| = 0 \ , \ea i.e for all $\vec a, \vec b$,
$|M_{\lambda}^A(\vec a)| = |M_{\lambda}^B(\vec b)| = 0$ for ``almost
all" $\lambda$ (except for a zero-measure subset of $\lambda$'s,
that could possibly depend on $\vec a, \vec b$).

\bigskip

\section*{Acknowledgements}

We thank A. Ekert and C. Simon for useful comments. C.B., N.B., and
N.G. aknowledge financial support from the EU project QAP (IST-FET
FP6-015848) and Swiss NCCR Quantum Photonics. A.L.L. and C.K.
acknowledge ASTAR for financial support under SERC grant No.
052~101~0043. This work is partly supported by the National Research
Foundation and Ministry of Education, Singapore.

Correspondence and requests for materials should be addressed to
C.B.

\end{document}


\title{Testing quantum correlations versus single-particle properties \\ within Leggett's model and beyond \\ $\,$ \\ SUPPLEMENTARY INFORMATION}
\author{Cyril Branciard, Nicolas Brunner, Nicolas Gisin}
\address{Group of Applied Physics, University of Geneva, Geneva, Switzerland}
\author{Christian Kurtsiefer, Antia Lamas-Linares, Alexander Ling, Valerio Scarani}
\address{Centre for Quantum Technologies / Physics Dept., National University of Singapore, Singapore}


\maketitle

\section*{SUPPLEMENTARY INFORMATION I: \\ TECHNICAL ASPECTS OF THE EXPERIMENT}

\label{section_exp}
\subsection{Pair source and detection}
The polarization-entangled photon pair source in our experimental
setup 
(Figure~2 of the article) is based on a non-collinear type-II
parametric down conversion process in a 2\,mm thick properly cut
Barium-beta-borate crystal, where photon pairs with almost
degenerate wavelengths around 702\,nm are collected \cite{pairsrc1}.
As a pump source (PS), we used an Argon ion laser with a center
wavelength of 351~nm and a power of 38\,mW. Longitudinal and partial
transverse walk-off compensation was established in the usual way
using a half wave plate and two 1\,mm thick BBO crystals (CC) of the
same cut as the conversion crystal \cite{pairsrc2}.

For the polarization measurements we used zero order quarter wave plates
($\lambda/4$) and polymer-film based polarization filters (PF) with a small
wedge error and an extinction ratio of better than 1:10,000. All four elements
could be oriented with motorized rotation stages with a reproducibility of
about 0.1$^\circ$.

Photo events were generated with
fiber-coupled passively quenched silicon avalanche photodiodes
(D$_{A,B}$). All photo events were recorded, and pairs identified
when photo events coincided in a time window of $\tau=15$\,ns.

A pair of interference filters (IF, center wavelength 702\,nm, full
transmission width 5\,nm at half maximum) was used for initial
alignment of the pair source, and left in the system to avoid a
change in beam orientations.

\subsection{Alignment and setup characterization}
As a first step in our alignment procedure, the single mode optical
fibers were adjusted for neutral polarization transport with
polarization controllers (FPC). Under stable temperature conditions,
we maintained an extinction ratio better than 1:1000 over half a
day. Then, the exact orientation $\theta_B$ of the polarizing filter
with respect to the first one was fixed to an accuracy of
0.2$^\circ$ by searching for a minimum of coincidence events, with
the quarter wave plates removed. Compensation crystals were adjusted
to maximize the visibility of polarization (anti-)correlations in
the $\pm45^\circ$ basis. At this point, we were reasonably confident
that we observe photon pairs in a good approximation of a singlet
polarization state. Typical visibilities we reached were
$V_{HV}\approx99\%$  and $V_{\pm45}\approx98\%$.

There is no direct evidence in the experimental result
for $L_3(\varphi)$ if the orientation of the quarter wave plates with
respect to the polarizing filters is accurate. We thus verified the
orientation of the first wave plate by symmetrization of the
polarization correlations with a fixed polarization filter in Bob's
arm under 0$^\circ$ linear polarization, and varying $\theta_B$
while observing coincidences. With this procedure, we were able to
find the symmetric position within $\Delta\gamma_A\approx0.2^\circ$.
The quarter wave plate in Bob's arm was oriented in a similar way.

To test the inequality 
(5) of the article, we used the following Poincar\'e vectors (see
Figure 
1): \ba \vec a_1 = \vec x \ , \quad \vec b_1 / \vec b_1' = \left(
\cos \frac{\varphi}{2}, \pm \sin \frac{\varphi}{2}, 0 \right),
\nonumber
\\ \vec a_2 = \vec y \ , \quad \vec b_2 / \vec b_2' = \left( 0, \cos
\frac{\varphi}{2}, \pm \sin \frac{\varphi}{2} \right), \nonumber
\\ \vec a_3 = \vec z \ , \quad \vec b_3 / \vec b_3' = \left(\pm \sin \frac{\varphi}{2}, 0, \cos
\frac{\varphi}{2} \right), \ea so that \ba \vec b_i - \vec b_i' = 2
\sin \frac{\varphi}{2} \vec e_i \, , \ \textrm{with} \ (\vec e_1,
\vec e_2, \vec e_3) = (\vec y, \vec z, \vec x). \ea

The $x$ axis in this notation corresponds to $\pm45^\circ$ linear,
the $y$ axis to circular, and the $z$ axis to horizontal/vertical
polarization, the latter coinciding with the natural basis of the
parametric down conversion process in the nonlinear optical crystal.

The correlation coefficients $C(\vec a,\vec b)$  for each of the six
settings for testing inequality~(5
) were
obtained by recording photo-detection coincidences $c$ during an
integration time $T$ for four combinations of orthogonal
polarizations by rotating the polarizers by additional 90$^\circ$
accordingly: \ba\label{eq_correlation_expression} C(\vec a, \vec
b)={c(\vec a,\vec b)+c(-\vec a,-\vec b) - c(-\vec a,\vec
  b)-c(\vec a,-\vec b) \over
  c(\vec a,\vec b)+c(-\vec a,-\vec b) + c(-\vec a,\vec b) +c(\vec
a,-\vec b)} \, . \ \ \ \ea This way, artefacts due to imbalanced
detector efficiencies were minimized.

\begin{figure}
\centerline{\epsfxsize=78mm \epsfbox{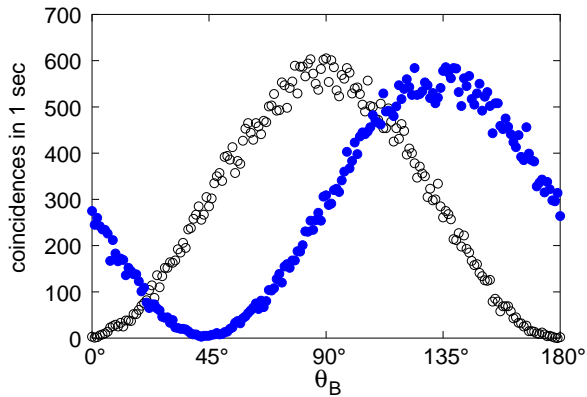}}
\caption{Polarization correlations
  in the $HV$ basis (open circles, $\theta_A=0^\circ$) and $\pm45^\circ$ linear
  basis (filled circles, $\theta_A=45^\circ$). The corresponding visibilities
  are $V_{HV}=98.9\pm0.8\%$ and $V_{\pm45^\circ}=97.8\pm0.8\%$.}
\label{fig_visis}
\end{figure}

In order to ensure the stability of the alignment, we recorded
polarization correlations in the $HV$ and $\pm45^\circ$ linear
polarization bases both before and after the acquisition of the
correlations for $L_3$ in the usual way (i.e., $\gamma_A=\theta_A$
fixed, $\gamma_B=\theta_B$ varies). Supplementary
Figure~\ref{fig_visis} shows these polarization correlations
directly before the main measurement. We extract visibilities of
$V_{HV}=98.9\pm0.8\%$ and $V_{\pm45^\circ}=97.8\pm0.8\%$,
respectively.

The uncertainties in all experimental values of $L_3$ are obtained
assuming Poissonian counting statistics for photodetection events,
independent variations for each setting, and the usual error
propagation through \ref{eq_correlation_expression} and (5).

To track potential drifts of the apparatus, the set shown in
Figure~3 of the article was recorded in two interleaving series with
a spacing of 5$^\circ$ for $\varphi$ each. The absence of a
modulation of $L_3(\varphi)$ with a $5^\circ$ period in the
presented graph indicates that the residual asymmetry can not be
attributed to drifts of our setup, but to a residual misalignment of
the quarter wave plates with respect to each other. From a
simulation with a noisy singlet state, the residual asymmetry seems
to be compatible with a misalignment on the order of $0.2^\circ$.

The polarization correlations after these measurements revealed
values of $V_{HV}=99.3\pm0.8\%$ and $V_{\pm45^\circ}=97.8\pm0.8\%$,
indicating again that the alignment of source, optical fibers and
measurement system has not drifted significantly over the
measurement time of about 9\,h.

\section*{SUPPLEMENTARY INFORMATION II: \\ OTHER LEGGETT-TYPE INEQUALITIES}
\label{app_other_ineqs}

Inequality 
(5) in the article is certainly the simplest one that allows to test
Leggett's model versus quantum mechanics. It involves only 3
settings for Alice and 6 for Bob. Note that on Alice's side, two
settings could actually be chosen to be the same: in our case, one
could for instance rotate the triplet $(\vec a_1,\vec b_1,\vec
b_1')$ around the $y$ axis, so that $\vec a_1 = \vec a_3 = \vec z$.
Thus, $2 \times 6$ settings are actually enough \footnote{Of course,
we do not claim that this number of settings ($2 \times 6$) is
minimal. We have found other inequalities, some with terms of the
form $|C(\vec a,\vec b) - C(\vec a,\vec b')|$, that use fewer
settings but which are less robust.}. In addition, inequality
(5) is stronger than the previously derived inequalities
\cite{leg03,Nature_Vienna,PRL_Vienna,PRL_Singap}. Consider for
instance the simplest inequalities of \cite{PRL_Vienna,PRL_Singap}:
in our formalism, they would be obtained by (i) choosing four
triplets, in two orthogonal planes, and (ii) setting
$\vec{a}_i=\vec{b}_i$. Now, the choice of the triplets was not
optimal, nor was the choice (ii), as the geometry of the problem
clearly shows.

Following our reasoning, one can derive even more robust
inequalities by increasing the number $N$ of (judiciously chosen)
triplets $(\vec a_i,\vec b_i,\vec b_i')$, thus defining $N$
directions $\vec{e}_i$ for $\vec b_i - \vec b'_i$. If $\frac{1}{N}
\sum_{i=1}^N |\vec{v} \cdot \vec{e}_i| \geq \xi_N
> 0$ holds for all $\vec v$, then one gets the Leggett-type
inequality
 \ba L_N(\varphi) & \leq & 2 - 2 \,
\xi_N \, |\sin \frac{\varphi}{2}| \ , \label{Leggett_ineq_N} \ea
with $L_N(\varphi) = \frac{1}{N} \sum_{i=1}^N |C(\vec a_i,\vec b_i)
+ C(\vec a_i,\vec b'_i)|$. This inequality is violated by the
singlet state, for which $L_N(\varphi)$ is still
equal to $L_{\Psi^-}(\varphi) = 2 |\cos\frac{\varphi}{2}|$. The
threshold visibility is now $V_{th}^{(N)} = \sqrt{1-\xi_N^2}$. The
next simple case after the one we studied is $N=4$: in this case the
four directions $\vec{e}_i$ should point to the four vertices of the
regular tetrahedron. One can then show that $\frac{1}{4}
\sum_{i=1}^4 |\vec{v} \cdot \vec{e}_i| \geq \frac{1}{\sqrt{6}}$
(i.e. $\xi_4 = \frac{1}{\sqrt{6}}$), and one gets an inequality that
tolerates a visibility of $V_{th}^{(4)} = \sqrt{\frac{5}{6}} \simeq
91.3 \%$. In the infinite (and experimentally non-testable) limit of
all directions $\vec e$ scanned by the vectors $\vec b_i - \vec
b'_i$, we would have $\xi_N \to \frac{1}{2}$,  and $V_{th}^{(N \to
\infty)} = \frac{\sqrt{3}}{2} \simeq 86.6 \%$.

\section*{SUPPLEMENTARY INFORMATION III: \\ A STRAIGHTFORWARD GENERALIZATION \\ OF LEGGETT'S MODEL}
\label{section_eta_leggett}

Here we come back to the generalization of Leggett's model, in which
we impose that the marginals should have the following form:
\ba \begin{array}{lcl} \moy{\alpha}_{\lambda} &=& \eta \ \vec{u} \cdot \vec{a} \ , \\
\moy{\beta}_{\lambda} &=& \eta \ \vec{v} \cdot \vec{b} \
,\end{array}& \mathrm{with} & 0 < \eta \leq 1 \ . \ea

This generalized model satisfies inequalities similar to
\ref{Leggett_ineq_N}: \ba L_N(\varphi) & \leq & 2 - 2 \, \eta \,
\xi_N \, |\sin \frac{\varphi}{2}| \ . \label{ineg_eta_leggett} \ea
For any value of $\eta > 0$, these inequalities are violated by the
quantum mechanical prediction $L_{\Psi^-}(\varphi)$, for a small
enough angle $\varphi$. Thus, such generalizations of Leggett's
model also fail to reproduce quantum mechanical predictions, and can
also be tested experimentally.

The required visibility to violate inequalities
\ref{ineg_eta_leggett} depends on $\eta$: $V_{th}^{(N)} =
\sqrt{1-(\eta \, \xi_N)^2}$. In the experimental run shown in Figure
3, we had $N = 3$, $\xi_3 = \frac{1}{3}$,
and an average visibility of $98.4\pm0.4\%$ . Therefore our data
allows us to falsify these generalized models for all $\eta
>0.528\pm0.07$. The statistical significance of the experimental excess of
$L_3$ with respect to the generalized Leggett bound increases
roughly linearly with $\eta$, from 3.65\,$\sigma$ for $\eta = 0.65$
up to 40.6\,$\sigma$ of the Leggett model bound
(5) for $\eta=1$.